# Dynamic Spatiotemporal Beams that Combine Two Independent and Controllable Orbital-Angular-Momenta Using Multiple Optical-Frequency-Comb Lines

**Zhe Zhao[1,*], Hao Song[1], Runzhou Zhang[1], Kai Pang[1], Cong Liu[1], Haoqian Song[1], Ahmed Almaiman[1], Karapet Manukyan[1], Huibin Zhou[1], Brittany Lynn[2], Robert W. Boyd[3,4], Moshe Tur[5], and Alan E. Willner[1,*]**

1. Department of Electrical Engineering, Univ. of Southern California, Los Angeles, CA 90089, USA
2. Space & Naval Warfare Systems Center, Pacific, San Diego, CA, 92152, USA
3. Department of Physics, University of Ottawa, Ottawa, ON, Canada
4. The Institute of Optics, University of Rochester, Rochester, New York 14627, USA
5. School of Electrical Engineering, Tel Aviv University, Ramat Aviv 69978, Israel

Corresponding emails: zhezhao@usc.edu, willner@usc.edu

Novel forms of beam generation and propagation based on structured light [1,2,3,4,5] and orbital angular momentum (OAM) [6,7,8,9,10,11,12] have gained significant interest over the past several years. Indeed, dynamic OAM can manifest at a given propagation distance in different forms [13], including: (1) a simple Gaussian-like beam "dot" [14,15,16] "revolves" around an offset central axis in time, and (2) a Laguerre-Gaussian ($LG_{\ell,p}$) beam with a helically "twisting" phase front [17,18] that "rotates" around its own central null in time. In this paper, we numerically generate dynamic spatiotemporal beams that combine these two forms of orbital-angular-momenta by coherently adding multiple frequency comb lines such that each carries a superposition of multiple $LG_{\ell,p}$ modes containing one $\ell$ value but multiple $p$ values. The generated beams can have different non-zero rotating $\ell$ values with high modal purities that exhibit both "rotation" and "revolution" in time at a given propagation distance. In our simulation results, we were able to control and vary several parameters, including the: (i) rotating $\ell$ value from +1 to +3 with modal purities of ~96%, (ii) revolving speed of 0.2-0.6 THz, (iii) beam waist of 0.15-0.5 mm, and (iv) revolving radius of 0.75-1.5 mm.

## Main

Structured light is a fascinating topic in that it can produce uniquely propagating beams of light [1,2,3,4,5]. One particularly interesting aspect is the ability of a light beam to carry orbital angular momentum [6,7,8,9,10,11,12]. One form of momentum is a simple Gaussian beam "dot" that can dynamically rotate in a circular fashion, illuminating a ring shape [14,15,16]; this OAM is similar to "revolution". A second form of momentum is a subset of Laguerre-Gaussian beams in which the phase front "twists" in the azimuthal direction as it propagates [17,18]. The amount of OAM ($\ell$) is the number of $2\pi$ azimuthal phase changes, and the beam rotates around its center with a ring-like vortex intensity profile (Fig. 1b1,c1). This second type of OAM is similar to "rotation". Indeed, the earth propagating around the sun exhibits both rotation around its "core" axis and revolution around an offset central "circular" axis [19].

These two manifestations of momentum can occur in space during propagation, but yet the beam will appear "static" at any given point of propagation distance [20,21,22,23,24]. Another added layer of complexity would be the generation and propagation of a "dynamic" spatiotemporal beam, such that the beam simultaneously revolves and rotates in the *x-y* plane at a given propagation distance *z*. Prior art has produced novel beams by combining not only different structured beams on the same frequency [20,21,22,23,24] but also different beams located at different frequencies [14,15,16,25,26,27,28]. Indeed, advances in optical frequency combs have enabled much novelty in many applications [29].

Specifically, it has been previously shown that a light beam can be created to exhibit unique ***dynamic*** features [14,15,16,25,26,27,28], including the following: (a) a Gaussian-like beam "dot" that exhibits dynamic circular ***"revolution"*** at a given propagation distance by combining multiple frequency lines in which each line carries a different $LG_{\ell,p}$ mode (different $\ell$ but same $p$) [14,15,16] (Fig. 1b2,c2); (b) a Gaussian-like beam "dot" formed from multiple Hermite-Gaussian modes, each with a different frequency, such that the dot can dynamically move ***"up-and-down"*** in a linear fashion at a given propagation distance [28]; (c) a light beam created by a pair of $LG_{\ell,p}$ modes with different $\ell$ and $p$ values at two different frequencies, such that it exhibits dynamic ***"rotation"*** around its center but no dynamic "***revolution***" around another axis at a given propagation distance [26]; and (d) a combination of multiple frequency lines, each of which carries one $LG_{\ell,p}$ mode with a different pair of indices $(\ell, p)$ and can produce a light beam that exhibits dynamic ***"rotation"*** around its center (azimuthal dimension) as well as ***"in-and-out"*** linear radial movement at a given propagation distance [27]. A laudable goal would be to produce a more sophisticated beam that can dynamically rotate and revolve at a given point in space.

In this paper, we numerically generate a revolving $LG_{\ell,0}$ beam combining two independent and controllable orbital-angular-momenta using multiple optical frequency comb lines, with each carrying a superposition of multiple $LG_{\ell,p}$ modes containing one unique $\ell$ value but multiple $p$ values (Fig. 1b3,c3). By varying the rotating $\ell$ value, revolving speed, revolving radius, and beam waist, we showed via simulation that we were able to control not only the spatiotemporal beam's helically "twisting" phase front as well as its dynamic, two-dimensional (2D) motion of *"rotation"* and *"revolution"* at a given propagation distance.

One example of such a beam propagation is an $LG_{\ell,p}$ beam with $p = 0$ rotating around its own central core while it also revolves around another offset central axis. We defined the revolving speed, $f_r$ revolutions/sec (or Hz), as the number of circles that a light beam revolves around an axis per second. If it is desired to generate such a spatiotemporal beam revolving at a speed of $f_r$ with a revolving radius of $R$, then its electrical field would be expressed as $E(x,y,z,t) \propto LG_{\ell,0}(x + R\cos(\omega_r t), y + R\sin(\omega_r t), z, \omega_0)\exp(-j\omega_0 t + j\varphi_0)$, where $\omega_r = 2\pi f_r$ is the revolving angular speed, $\omega_0 = 2\pi f_0$ is the angular frequency of the revolving beam, $\varphi_0$ is the initial phase delay, $(x + R\cos(\omega_r t), y + R\sin(\omega_r t), z, t)$ is the coordinate transformation of $(x, y, z, t)$ in a reference frame rotating at the transverse plane, and $LG_{\ell,0}(x, y, z, \omega_0)$ is the electrical field of a stationary $LG_{\ell,0}$ beam. Such a spatiotemporal beam can also be described as a superposition of multiple frequency comb lines in which each line carries a unique spatial pattern. The electrical field can be written in a form that is modified from [15]:

$$E(x,y,z,t) = \sum_{i,k} C_{i,k} LG_{\ell_i,p_i}(x,y,z,\omega_k)\exp(-j\omega_k t + j\varphi_k) \quad (1)$$

where $LG_{\ell_i,p_i}(x,y,z,\omega_i)$ ($i = 1, 2, \dots, n$) is the electrical field of the $LG_{\ell_i,p_i}$ beam; $C_{i,k}$ is the complex coefficient weight; $\omega_k$ is the angular frequency; $\varphi_k$ is the initial phase delay; and $(x, y, z, t)$ are the coordinate and time, respectively. By choosing the angular frequency, spatial pattern, and complex coefficients, a "structured" beam with unique dynamic spatiotemporal properties can be generated.

By way of example, we simulated the dynamic motion of an $LG_{3,0}$ beam (beam waist $w_0 = 0.3$ mm center frequency $f_0 = 193.5$ THz) revolving around an axis with a radius of $R = 0.75$ mm at a speed of $f_r = 0.2$ THz. We used 61 frequency comb lines with a frequency spacing $\Delta f$ of 0.2 THz and the same initial phase delay of $\varphi = 0$. Each line is a superposition of multiple $LG_{\ell,p}$ modes containing one $\ell$ value but multiple $p$ values, where $\ell$ is proportional to $f$, and $p$ varies from 0 to 25. The electrical field can be represented by $\sum_{\ell=-30}^{30} \sum_{p=0}^{25} C_{\ell,p} LG_{\ell,p}(x, y, 0, \omega_\ell) \exp(-j\omega_\ell t)$ at distance $z = 0$, where $\omega_\ell = 2\pi(f_0 + \ell\Delta f)$ is linearly correlated with the azimuthal mode index $\ell$, and the frequency line at $\omega_\ell$ carries a superposition of spatial patterns $\sum_{p=0}^{25} C_{\ell,p} LG_{\ell,p}(x, y, 0, \omega_\ell)$.

We characterized the beam's spatial spectrum using the amplitude and phase of the complex coefficients $C_{\ell,p}$ for each $LG_{\ell,p}$ mode (Fig. 2b). Moreover, we mapped the spatial spectrum onto the frequency spectrum based on the linear correlation between the mode index $\ell$ and the angular frequency $\omega_\ell$ (Fig. 2a). Specifically, we calculated the total power on each frequency comb line using the total power of the superposition of $\sum_{p=0}^{25} C_{\ell,p} LG_{\ell,p}(x, y, 0, \omega_\ell)$ (see supplementary Fig. S1 for the spatial patterns on selected frequency lines). Additionally, the phase front and envelope (equi-amplitude surface) structures of such a beam were simulated, in which the mode purity of the generated revolving $LG_{3,0}$ beam was ~ 96% (see Fig. 2d and Methods). As shown in Fig. 2e, the dynamic helical phase front and envelope indicated that the beam exhibited both dynamic *"rotation"* and *"revolution"* at a given distance.

The light beam exhibits both dynamic *"rotation"* and *"revolution"* due to the coherent interference among the multiple frequency comb lines, each of which carries a superposition of multiple $LG_{\ell,p}$ modes. First, such a coherent interference results in a "twisting" phase front of $\exp(j\ell\theta)$ in a circle around an intensity null, leading to a Poynting vector with a non-zero azimuthal component. Because a light beam propagates along the Poynting vector in free space, it exhibits dynamic *"rotation"* at a given propagation distance [18]. Second, the interferogram of multiple $LG_{\ell,p}$ modes with different $\ell$ values on a single frequency is an azimuthal mode, such that introducing a constant relative phase delay of $\Delta\varphi$ between neighbouring $LG_{\ell,p}$ modes (*i.e.*, $\Delta\ell = 1$) will rotate the azimuthal mode by an angle of $\Delta\varphi$ [30]. Because the generated beam can be regarded as "dynamic" interferogram of multiple $LG_{\ell,p}$ modes with different $\ell$ values located on multiple frequencies, where $\omega_\ell = 2\pi(f_0 + \ell\Delta f)$, there will be a time-variant relative phase delay of $2\pi\Delta ft$ between the so-called neighbouring $LG_{\ell,p}$ modes. Therefore, the light beam also exhibits dynamic *"revolution"* at a given propagation distance.

We showed that these two momenta can be independently and separately controlled by tuning the revolving speed and the rotating $\ell$ values of different revolving $LG_{\ell,p}$ beams, which are associated with the *"revolution"* and *"rotation"*, respectively (Fig. 3). We investigated the cases for a revolving $LG_{\ell,0}$ beam (i) revolving clockwise at a speed of 0.2 THz and carrying a rotating $\ell$ value varied from 1 to 3 (Fig. 3a1,a2,a3), or (ii) carrying the same rotating $\ell = 3$ value and revolving at a speed varied from 0.2 THz to 0.6 THz (Fig. 3b1,b2,b3). Two main conclusions arose. First, the rotating $\ell$ value of the revolving $LG_{\ell,0}$ beam can be controlled by changing the spatial $LG_{\ell,p}$ mode distribution on each frequency line (see supplementary Fig. S2 for details). Second, the $LG_{\ell,0}$ beam revolves at a speed equal to the frequency spacing $\Delta f$. Moreover, the $LG_{\ell,0}$ beams can be tuned to revolve counter-clockwise if we flip the sign of the $\ell$ value of each $LG_{\ell,p}$ mode carried by the frequency comb lines [14,15].

Furthermore, we investigated the quality of the dynamic spatiotemporal beam as it relates to the frequency comb spectrum. Figures 4a,b,c show the relationship between the power distribution on light beams with different rotating $\ell$ values and the number of selected frequency comb lines. Figure 4b shows that when the number of comb lines is selected to be < 10, the power coupling to the light beams with the rotating $\ell \neq 3$ value is > -5 dB and the mode purity of the generated revolving $LG_{3,0}$ beam is < 50%; while when the number of comb lines is > 40, the power coupling is < -12 dB and the mode purity is > 95%. We can see from Fig. 4c that ~ 30 frequency lines are needed to generate a revolving $LG_{\ell,0}$ beam, where $\ell = 0, 1, 2, 3$, with a mode purity of > 90%. For the cases where a limited number of frequency lines, such as 20, were used, the mode purity of the generated revolving $LG_{\ell,0}$ beam was higher for smaller rotating $\ell$ values. Figures 4d,e,f show the number of frequency comb lines within the 10-dB bandwidth of the frequency spectra for generating revolving $LG_{\ell,0}$ beams with different revolving radii or beam waists. The simulation results show that a larger number of frequency comb lines would generate a revolving $LG_{\ell,0}$ beam with a (i) larger revolving radius, (ii) smaller beam waist, or (iii) higher rotating $\ell$ value. These relationships can be understood by using a Fourier transformation: Looking at the "dynamic" azimuthal mode (the generated spatiotemporal beam) at a given time, it can be described a superposition of multiple $LG_{\ell,p}$ modes with different azimuthal index $\ell$ values [30]. As the light beam's (i) revolving radius increases, (ii) beam waist decreases, or (iii) rotating $\ell$ value increases, the azimuthal mode will have a smaller azimuthal feature; thus the number of comb lines increases after applying a Fourier transformation from the azimuthal spatial domain to the frequency domain [30].

**Methods**

**Additional simulation details**

The electrical field of an $LG_{\ell,p}$ mode can be described by [18]:

$$LG_{\ell,p}(r,\theta,z,\omega) = \frac{C_{\ell,p}^{LG}}{w(z)} \left(\frac{r\sqrt{2}}{w(z)}\right)^{|\ell|} \exp\left(-\frac{r^2}{w^2(z)}\right) L_p^{|\ell|}\left(\frac{2r^2}{w^2(z)}\right) \exp\left(-i\left(k\frac{r^2}{2R(z,\omega)} + \ell\theta + kz - \psi(z,\omega)\right)\right)$$

(M1)

where $L_p^{|\ell|}$ are the generalized Laguerre polynomials, and $C_{\ell,p}^{LG}$ are the required normalization constants, $w(z)$ is the beam radius, and $R(z,\omega) = z(1 + (z_R(\omega)/z)^2)$, where $z_R(\omega)$ is the Rayleigh range, $k$ is the wave number; and $\psi(z)$ is the Gouy phase and equals $(|\ell| + 2p + 1)\arctan(z/z_R(\omega))$. The parameters $(r,\theta,z,\omega)$ have the same definitions as in the letter.

We numerically generated the spatiotemporal beam in three steps: (i) we first calculated the complex spatial mode distribution of a stationary $LG_{\ell,0}$ beam centered at $(x,y) = (-R,0)$ by decomposing its electrical field into an $LG_{\ell,p}$ mode basis centered at $(x, y) = (0,0)$. The frequency is $f_0$, the revolving radius is $R$, $z = 0$, and $t = 0$; (ii) we then coherently combined all the spatial modes with the same $\ell$ value but different $p$ values to obtain the spatial pattern of the frequency line at $\omega_\ell = 2\pi(f_0 + \ell\Delta f)$, where $\Delta f$ is the revolving speed; (iii) we calculated the electrical field of the revolving $LG_{\ell,0}$ beam by coherently combining the electrical fields of all the frequency comb lines. We only considered the cases in which the frequency separation $\Delta f$ is a constant and the center frequency is 193.5 THz. In our simulation model, there are $500 \times 500$ pixels with a 6-μm pixel size in the $(x, y)$ plane, and 400 pixels with a 12.5-fs pixel size in time.

**Mode purity calculation**

Considering that the observed intensity and phase profiles of the generated revolving $LG_{\ell,0}$ beams remain relatively invariant if an observer moves dynamically with the revolving beams (Fig. 2e), we calculated the mode purity as the normalized power weight coefficients of the generated spatiotemporal beam at time $t = 0$ and distance $z = 0$ using $|C_\ell|^2 = |\iint E_1(x,y)E_2^*(x,y)dxdy|^2$ [10], where $E_1(x,y)$ is the generated electrical field of the generated spatiotemporal beam, and $E_2(x,y)$ is the electrical field of a stationary pure $LG_{\ell,0}$ beam with its center overlaps with the generated beam. Here, the calculated mode purity represents the ratio between the power on the spatiotemporal beam with the desired rotating $\ell$ value and the total power of the generated beam.

**Acknowledgements**

This work is supported by Vannevar Bush Faculty Fellowship sponsored by the Basic Research Office of the Assistant Secretary of Defense (ASD) for Research and Engineering (R&E) and funded by the Office of Naval Research (ONR) (N00014-16-1-2813).

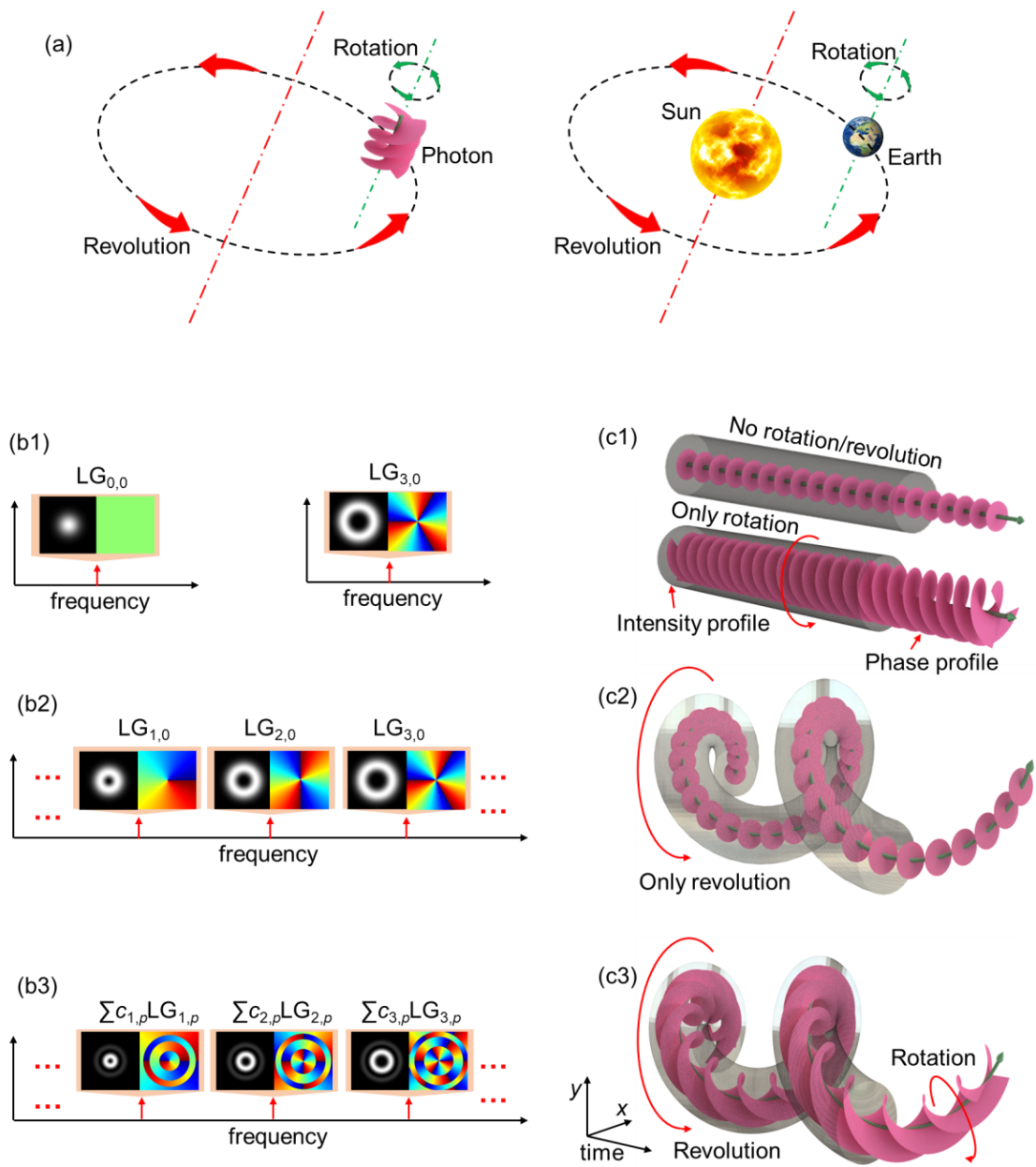

**Figure 1. The concept of generating a spatiotemporal beam exhibiting both rotation and revolution.** (a) Illustration of a light beam that dynamically rotates around its central core and revolves around another offset central axis; this is analogous to the earth orbiting around the sun, exhibiting both rotation around its "core" axis and revolution around the sun. (b1)-(b3) The concept of using frequency comb lines, in which each carries a different spatial mode distribution, to generate light beams that exhibit (c1) no rotation/revolution ($LG_{0,0}$ beam shown at the top), only rotation ($LG_{3,0}$ beam shown in the bottom), (c2) only revolution (revolving

Gaussian beam), and (c3) both rotation and revolution (revolving LG$_{3,0}$ beam). The time scales for rotation and revolution are differently scaled in (c1-3) in order to clearly illustrate the concept of rotation and revolution.

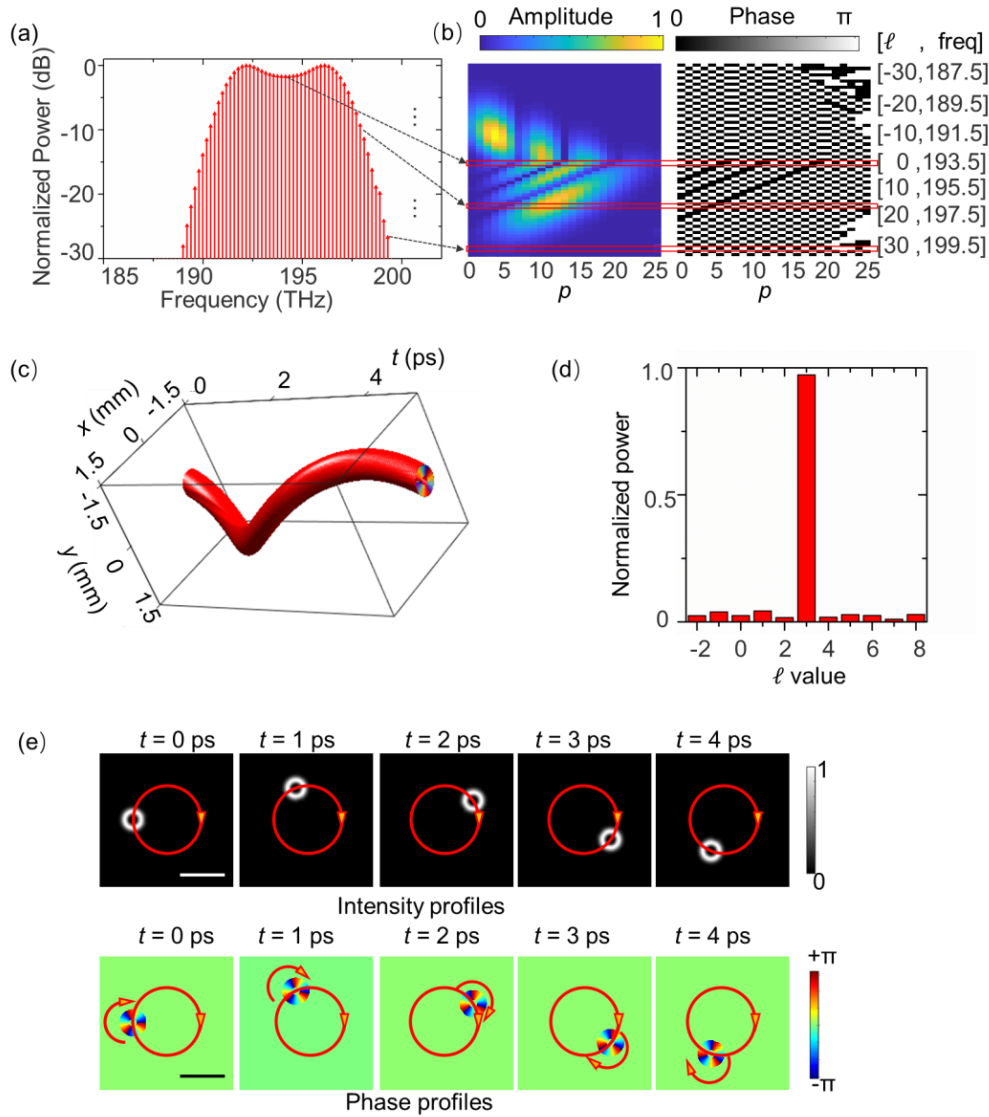

**Figure 2. Simulation results of a dynamic spatiotemporal beam exhibiting both rotation and revolution at a given propagation distance.** (a) frequency spectrum; (b) spatial LG$_{\ell,p}$ mode distribution, namely the amplitude and phase of the complex coefficients $C_{\ell,p}$ of all the LG$_{\ell,p}$ modes used for superposition; (c) the envelope structure (*i.e.*, the iso-surfaces with an amplitude of 1/10 of the peak value), where the cap represents the helically "twisting" phase front; (d) the power distribution on light beams with different rotating $\ell$ values for generating a revolving LG$_{3,0}$ beam, and (e) dynamically rotating and revolving intensity/phase profiles of

the generated $LG_{3,0}$ beam revolving at 0.2 THz. Scale bar, 1 mm. The spatiotemporal beam consists of multiple frequency comb lines, in which each line is a superposition of multiple $LG_{\ell,p}$ modes (same beam waist of 0.3 mm) with one $\ell$ value but multiple $p$ values. The dynamic helical wavefront and envelope indicates that the beam not only rotates around its central core but also revolves around another "circular" center axis 0.75 mm away from the rotating axis.

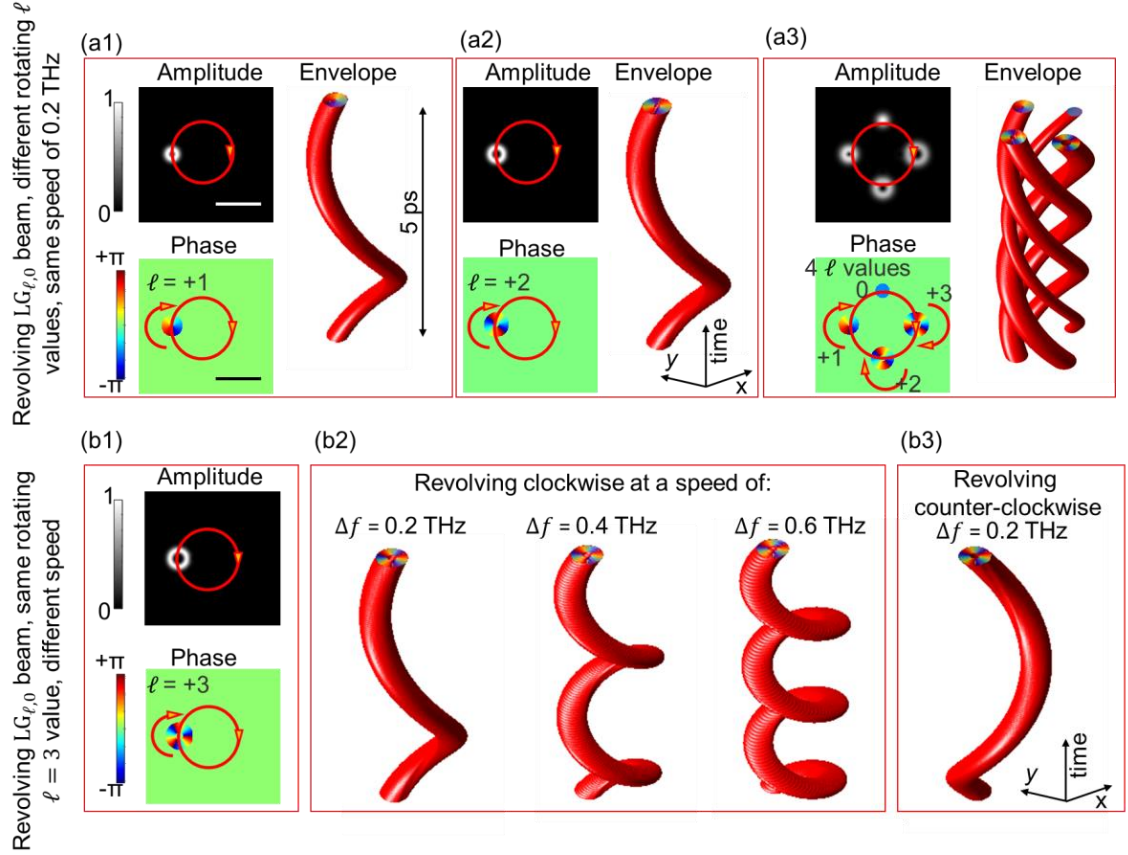

**Figure 3. Independent control of two momenta of the spatiotemporal beams**. (a1-3) The 2D amplitude and phase profiles at time $t = 0$, and the envelope structures in $(x, y, t)$ coordinates of the generated revolving $LG_{\ell,0}$ beams with different rotating $\ell$ values but the same revolving speed of 0.2 THz. Scale bar, 1 mm. The phase fronts are $\exp(j2\theta)$ and $\exp(j4\theta)$ in one circle around the beams' center intensity nulls in (a1,2), respectively; (a3) is a single beam combining an array of four revolving $LG_{\ell,0}$ beams, where $\ell = 0, 1, 2, 3$. See supplementary Fig. S2 for details of the spatial $LG_{\ell,p}$ mode distributions used for superposition. (b1) The corresponding profiles/structures of the generated revolving $LG_{\ell,0}$ beams with the same $\ell = 3$ value but a different revolving speed. (b2) Examples of a $LG_{3,0}$ beam revolving clockwise at different speeds from 0.2 THz to 0.6 THz; and (b3) an $LG_{3,0}$ beam revolving counter-clockwise at a speed of 0.2 THz. Except for the varied parameters and the spatial/frequency spectra, all the other parameters are the same as those in Fig. 2.

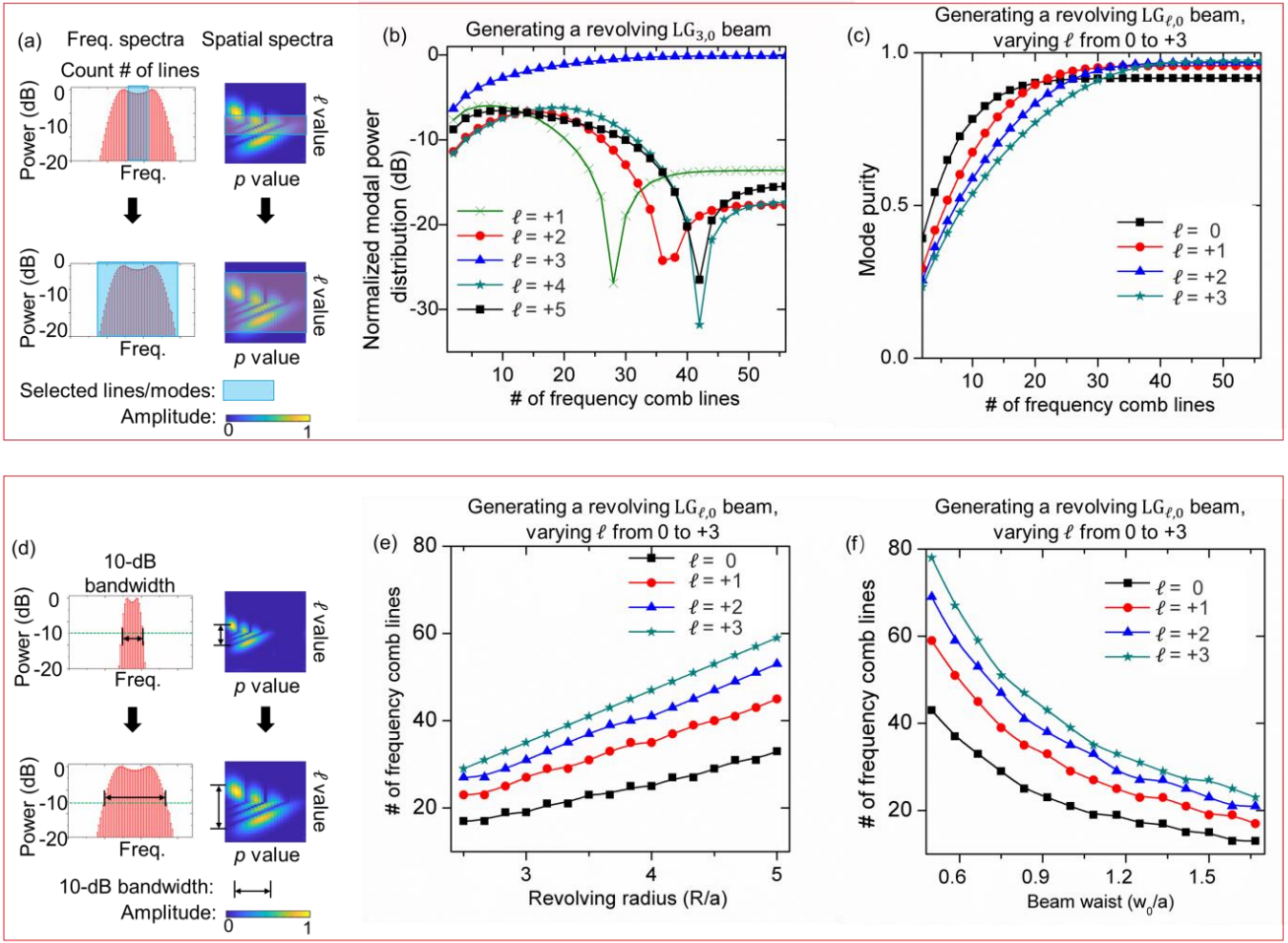

**Figure 4. The relationship between the quality of the revolving $LG_{\ell,0}$ beams and the frequency comb spectrum.** (a) We first calculated the spatiotemporal spectra of the revolving beam using the approach described in Methods, and then we selected a certain number of frequency lines/modes from the calculated spectra for power/mode purity calculation in (b) and (c). (b) The power distribution on light beams with different rotating $\ell$ values for generating a revolving $LG_{3,0}$ beam, when the number of selected frequency comb lines is varied. The blue curve represents the mode purity of the spatiotemporal beam, and the differences between the blue curve with the other curves represent the power coupling from the revolving $LG_{3,0}$ beams to the other light beams with different rotating $\ell$ values. (d) We calculated the spatiotemporal

spectra of different revolving beams and counted the number of frequency lines in the 10-dB bandwidth of the frequency spectra; (e) and (f) show the number of frequency comb lines in the 10-dB bandwidth for generating a revolving $LG_{\ell,0}$ beam with (i) the same beam waist of $w_0 = 0.2$ mm and a revolving radius of $R$ varied from 0.75 mm to 1.5 mm; and (ii) the same revolving radius of $R = 1.5$ mm and a beam waist of $w_0$ varied from 0.15 mm to 0.5 mm, respectively. Here, a = 0.3 mm is the beam waist of the $LG_{\ell,p}$ modes as the basis set for superposition.

# Supplementary: Dynamic Spatiotemporal Beams that Combine Two Independent and Controllable Orbital-Angular-Momenta Using Multiple Optical-Frequency-Comb Lines

**Zhe Zhao[1,*], Hao Song[1], Runzhou Zhang[1], Kai Pang[1], Cong Liu[1], Haoqian Song[1], Ahmed Almaiman[1], Karapet Manukyan[1], Huibin Zhou[1], Brittany Lynn[2], Robert W. Boyd[3,4], Moshe Tur[5], and Alan E. Willner[1,*]**

1. Department of Electrical Engineering, Univ. of Southern California, Los Angeles, CA 90089, USA
2. Space & Naval Warfare Systems Center, Pacific, San Diego, CA, 92152, USA
3. Department of Physics, University of Ottawa, Ottawa, ON, Canada
4. The Institute of Optics, University of Rochester, Rochester, New York 14627, USA
5. School of Electrical Engineering, Tel Aviv University, Ramat Aviv 69978, Israel

*Corresponding emails: zhezhao@usc.edu, willner@usc.edu*

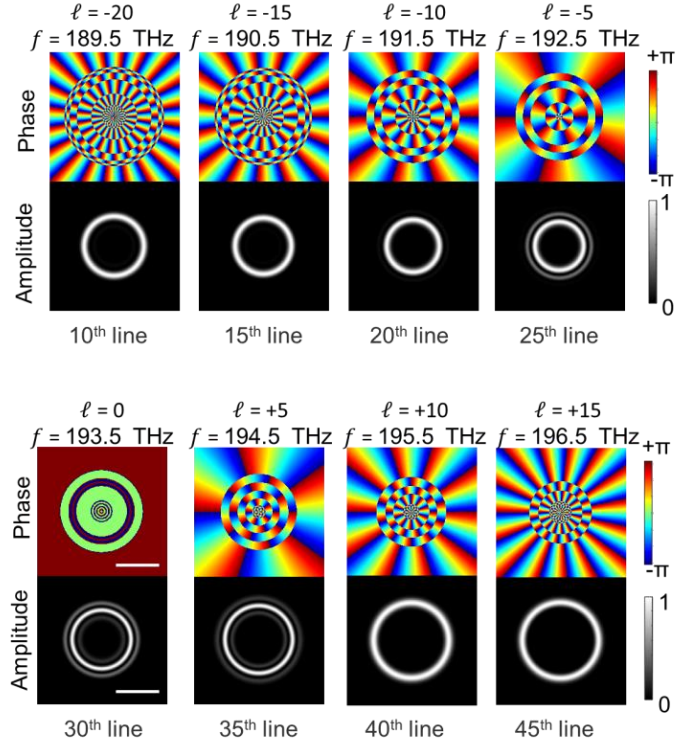

Figure S1. The spatial patterns on selected frequency lines for generating the revolving $LG_{3,0}$ beam in Fig. 2. Each frequency comb line carries a superposition of multiple $LG_{\ell,p}$ modes with one $\ell$ value but multiple $p$ values.

As shown in Fig. S1, we calculated the electrical fields of the selected frequency lines at distance $z=0$ and time $t=0$ by using $\sum_{p=0}^{25} C_{\ell,p} LG_{\ell,p}(x,y,0,\omega_\ell)$ for generating the spatiotemporal beam in Fig. 2. As expected, each frequency line has a unique "twisting" phase profile. The azimuthal $\ell$ values carried by the spatial pattern located on the $10^{th}$ to $45^{th}$ frequency lines are -20, -15, -10, -5, 0, +5, +10, and +15, respectively, which ensures that the $\ell$ value is proportional to $f$. Moreover, each frequency comb line carries a superposition of $LG_{\ell,p}$ with multiple $p$ values, which results in a unique radial intensity and phase distribution along the radial direction. We can see that some frequency comb lines have a multiple-concentric-ring intensity profile, and that the phase changes by $\pi$ between neighboring rings along the radial direction.

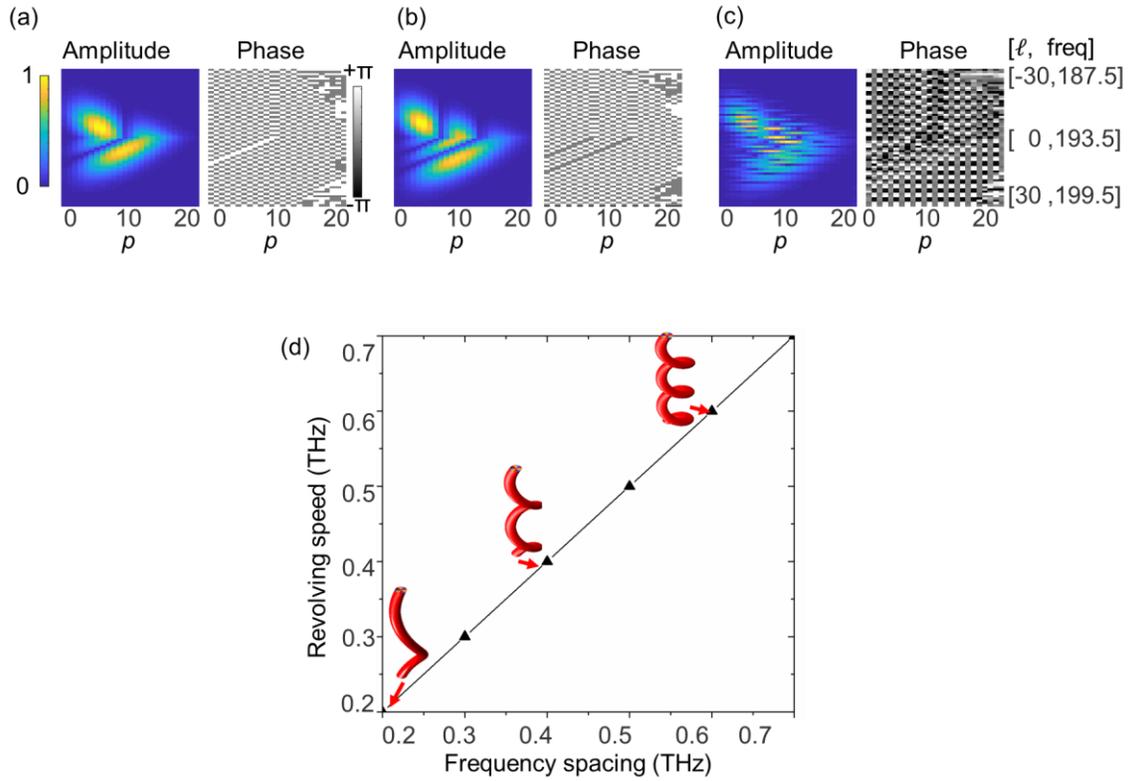

Figure S2. (a-c) The spatial $LG_{\ell,p}$ mode distributions, namely the amplitude and phase of the complex coefficients $C_{\ell,p}$ of all the $LG_{\ell,p}$ modes, for generating the three different revolving light beams shown in Fig. 3a. For generating revolving (a) $LG_{1,0}$ and (b) $LG_{2,0}$ beams, the amplitude matrices of the spatial $LG_{\ell,p}$ mode distribution are split into two and three main parts, respectively. (c) Spatial spectrum for generating an array of four revolving $LG_{\ell,p}$ beams with $\ell = 0, 1, 2,$ and 3. (d) The revolving speed of such a revolving $LG_{3,0}$ beam is equal to the frequency spacing.

Here, we first simulated the spatial mode distributions for generating revolving $LG_{1,0}$ and $LG_{2,0}$ beams and an array of four revolving $LG_{\ell,0}$ beams. Figure S2 shows that in order to change the rotating $\ell$ value of the generated revolving $LG_{\ell,0}$ beam, we needed to change both the amplitude and phase of the complex coefficients $C_{\ell,p}$ of the spatial $LG_{\ell,p}$ mode distribution used for superposition. For generating a revolving $LG_{\ell,0}$ beam with a non-zero rotating $\ell$ value, the spatial $LG_{\ell,p}$ mode distribution resembles a "heart" shape, which is then split into a number of $\ell$ main parts. As in the case of generating an array of revolving $LG_{\ell,0}$ beams with multiple $\ell$ values, there are several horizontal "noisy" lines in the "heart" shape, and most of its power is limited in a smaller number of $LG_{\ell,p}$ modes (Fig. S2c). We also showed the dependence of the revolving speed of the generated revolving $LG_{3,0}$ beam on the frequency spacing $\Delta f$, as shown in Fig. S2d. The revolving speed was tuned from 0.2 THz to 0.7 THz, which is equal to the frequency spacing $\Delta f$.

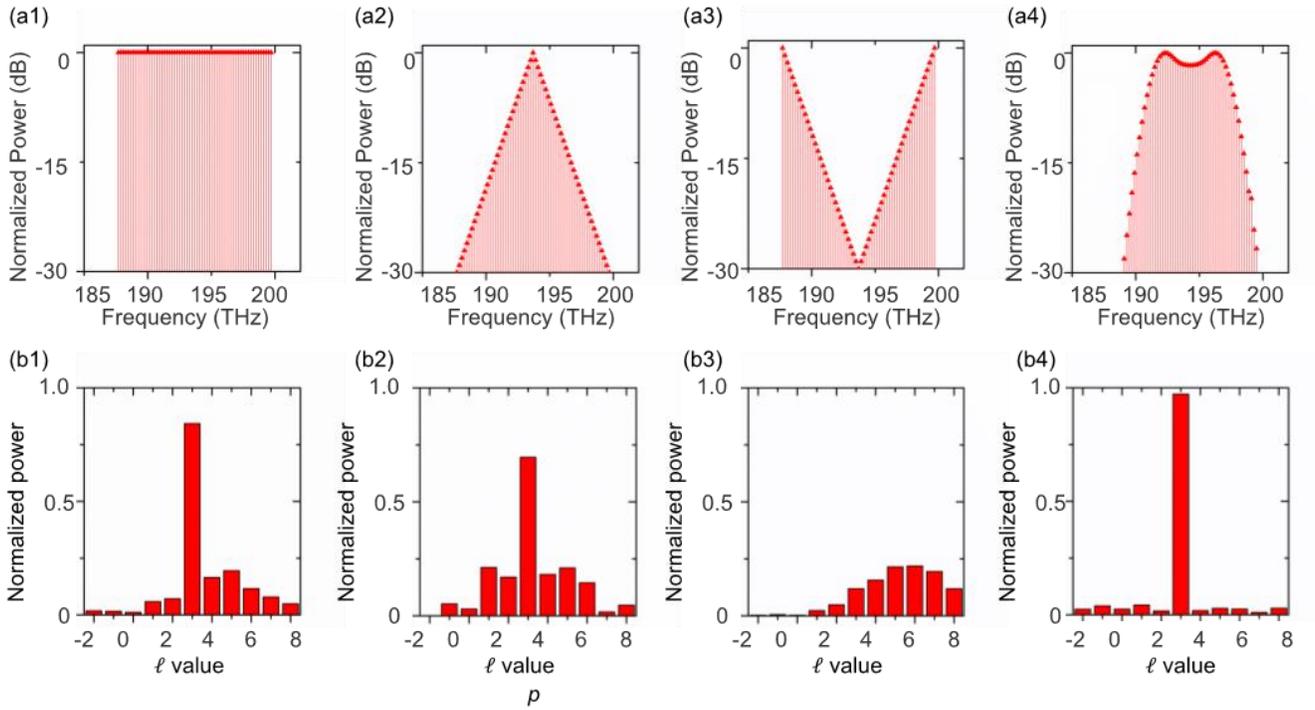

Figure S3. Effect of the frequency spectrum shaping on the normalized power distributions of light beams with different rotating $\ell$ values; all these frequency comb lines are used for generating a revolving $LG_{3,0}$ beam. (a4) is the same one in Fig. 2 in the letter.

Here, we characterized the quality of the generated beams by analyzing the mode purity as it relates to different frequency spectrum shapes (Fig. S3). We still calculated the mode purity as the normalized power weight coefficients of the generated vortices at time $t = 0$ and distance $z = 0$. Each comb line carries the same spatial pattern as shown in Fig. 2b in the letter (see selected patterns in Fig. S1), but the total power of each frequency line is varied. This shows that changing the frequency spectrum shape will affect the mode purity if the generated spatiotemporal beam. The mode purity of the beam with the shape shown Fig. S3a4 is ~ 96%. However, it decreases to ~ 84%, 70%, and 12% for the shapes in Fig. S3a1-a3, respectively. The third shape has the lowest mode purity because it filters most of the power of the $LG_{\ell,p}$ modes, leading to a higher spatial phase distortion.